\pdfoutput=1
\documentclass[fleqn,usenatbib]{mnras}

\usepackage{newtxtext,newtxmath}
\usepackage[T1]{fontenc}

\DeclareRobustCommand{\VAN}[3]{#2}
\let\VANthebibliography\thebibliography
\def\thebibliography{\DeclareRobustCommand{\VAN}[3]{##3}\VANthebibliography}

\usepackage{graphicx}	% Including figure files
\usepackage{amsmath}	% Advanced maths commands
\usepackage{xcolor}

\title[Connecting Cores to the MBHB Population]{Connecting Core Galaxy Properties to the Massive Black Hole Binary Population}

\author[Harris \& Gültekin]{
CJ Harris,$^{1}$\thanks{E-mail: cordellh@umich.edu (CJH)}
Kayhan Gültekin$^{1}$
\\
$^{1}$Department of Astronomy, University of Michigan, 1085 S University Ave, Ann Arbor, MI 48109
}

\date{Accepted 2023 October 31. Received 2023 October 04; in original form 2023 May 01}

\pubyear{2023}

\begin{document}
\label{firstpage}
\pagerange{\pageref{firstpage}--\pageref{lastpage}}
\maketitle

\begin{abstract}
We investigate how the properties of massive black hole binaries influence the observed properties of core galaxies. We compare the observed trend in stellar mass deficit as a function of total stellar mass in the core galaxy with predicted trends in IllustrisTNG. We calculate mass deficits in simulated galaxies by applying sub-grid, post-processing physics based on the results of literature N-body experiments. We find the median value of the posterior distribution for the minimum binary mass ratio capable of creating a core is 0.7. For the gas mass fraction above which a core is erased we find a median value of 0.6. Thus low mass ratio binaries do not contribute to core formation and gas-rich mergers must lead to star formation within the nucleus, ultimately erasing the core. Such constraints have important implications for the overall massive black hole binary population, black hole--galaxy co-evolution, and the origin of the gravitational wave background.
\end{abstract}


%%%%%%%%%%%%%%%%%%%%%%%%%%%%%%%%%%%%%%%%%%%%%%%%%%

%%%%%%%%%%%%%%%%% BODY OF PAPER %%%%%%%%%%%%%%%%%%

\section{Introduction}

\begin{figure}
	\includegraphics[width=\columnwidth]{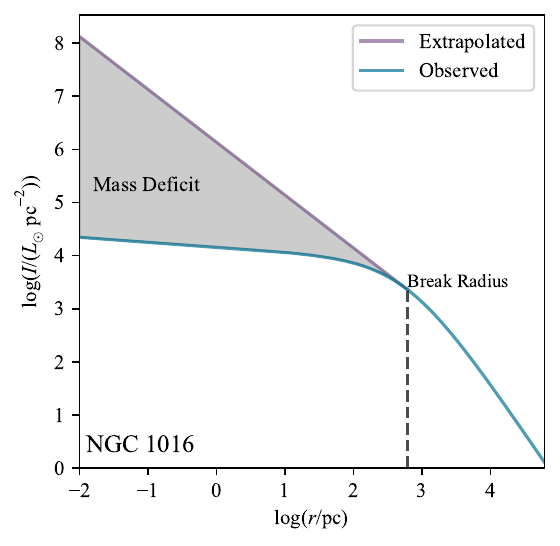}
    \caption{Stellar mass deficits in core galaxies may be calculated if the surface brightness profile is known and a model of the extrapolated light distribution is created. The light deficit is the integrated difference between the two profiles, which can be converted into a stellar mass deficit by multiplication of a stellar mass-to-light ratio in the appropriate band. The “break radius” indicates the distance from the galaxy center where the profile transitions from the shallow inner region to the steep outer region, representing the size of the core. Based on the profile of NGC 1016.}
    \label{fig:MassDefEx}
\end{figure}

The most massive elliptical galaxies have low-density centers $\lesssim$ 1 kpc in size \citep{2007ApJ...664..226L}, manifesting observationally as shallow inner surface brightness profiles compared to less massive ellipticals and to spiral galaxy classical bulges. The inner profile is significantly flatter than the steep, inward extrapolation of the outer profile (see Figure \ref{fig:MassDefEx}). Systems for which this is true are called “core” galaxies, while galaxies that have a steep inner profile are called ``power-laws''. Core galaxies are likely to have significantly different evolution, relative to other ellipticals, as they tend to be more spherical, have boxier isophotes, and rotate more slowly \citep{2012ApJ...759...64L, 2013MNRAS.433.2812K}.

It is thought that massive black hole binaries (MBHBs) are responsible for the formation of cores. Binary black holes are sources of gravitational waves and the precursor to black hole mergers, which is a channel of black hole growth. The connection between supermassive black holes (SMBHs) and their host galaxies
is important for understanding their co-evolution, including BH scaling relations and active galactic nuclei (AGN) feedback.

A number of observational facts about ETGs has lead to a theoretical picture of how galaxy cores are formed through MBHB activity. Most massive early-type galaxies (ETGs) like those in question have undergone at least one major merger since $z \sim 1$ \citep{2006ApJ...652..270B}. Between 50--70\% of the mass growth of ETGs has occurred in the same time-frame as a result of merger events, with about half of the contribution from major mergers \citep{2012A&A...548A...7L}. For galaxies with stellar masses $\gtrsim 10^{11}\ M\mathrm{_\odot}$ major mergers play a dominant role in mass assembly since $z \sim 1$, with 30\% of those mergers being gas-poor \citep{2009ApJ...697.1369B}. Accretion of less massive ETGs (i.e., minor mergers) also contributes to mass assembly \citep{2006ApJ...636L..81N}. Concerning rotation, the gas content of remnant galaxies plays a crucial role \citep{2022ApJ...925..168Y,2020MNRAS.494.5652W}, with gas-poor mergers tending to reduce the specific angular momentum of the stellar component by an average of $\sim30\%$, while gas-rich mergers are shown to increase angular momentum by $\sim10\%$ \citep{2018MNRAS.473.4956L}.

Given these properties, the formation of cores is best explained by stellar scattering in the galactic nucleus by MBHBs \citep{1980Natur.287..307B, 1997AJ....114.1771F} in the aftermath of gas-poor, elliptical-elliptical merger events. Once the progenitor galaxies merge, their respective SMBHs are initially unbound, eventually coupling near the remnant galaxy's nucleus via dynamical friction (DF). Once the black holes form a binary, they scatter neighboring stars, overcome depletion of the loss-cone, and begin to emit gravitational waves before coalescing. Thus, to successfully scour out a core, the secondary black hole must join the primary at the bottom of the remnant potential well. 

Prescriptions based on previous analyses \citep[e.g.,][]{2022MNRAS.510..531C} estimate the efficiency of DF as a function of the mass of the secondary, the central velocity dispersion of the remnant, the distance from the galactic center, the number of nuclear stars present, and the gas mass fraction. The DF timescale increases for larger merger remnants, since the secondary black hole must travel further to reach the primary. A high stellar velocity dispersion also leads to a longer timescale, as only stars moving slower than the secondary can contribute to friction. The DF timescale decreases for more massive secondaries and for a larger gas content. Once the binary is formed, stellar scattering ensues via three-body interactions. Scattering will tend to eject stars from the nucleus at the expense of the black hole binary's orbital energy, shrinking the separation between the black holes. The result is a low density core which is likely to persist for a long period, unless the gas content in the remnant is high enough to trigger star formation in the galactic center, which has the effect of erasing the core \citep{1997AJ....114.1771F}.

If the core is not erased, then it will appear in observations as a luminosity deficit in the galaxy center, which can be converted into a mass deficit given a mass-to-light ratio. N-body experiments have suggested that the mass in ejected stars is proportional to the mass of the central black hole and the number of major mergers, with between one and three major mergers ejecting a mass in stars consistent with observed mass deficits \citep{1996NewA....1...35Q,2001ApJ...563...34M, 2006ApJ...648..976M}. Additionally, MBHB N-body experiments predict a dearth of radial stellar orbits in the nucleus of core galaxies as such orbits have small pericenters, allowing for strong MBHB interactions that kick the stars out of the center.  This prediction is seen in detailed orbital modeling of elliptical galaxies with high-spatial resolution spectroscopy \citep{2014ApJ...782...39T}.

Despite the observational evidence that galaxy cores are formed by MBHBs, there are few confirmed binaries \citep[e.g.,][]{2006ApJ...646...49R, 2012ApJS..201...23E, 2015MNRAS.453.1562G, 2015ApJS..221....7R, 2017MNRAS.468.1683R, 2016MNRAS.463.2145C}, though it is thought they should exist in large number. One line of evidence is the population of dual AGN, where widely separated, gravitationally unbound pairs are known \citep{2012AJ....143..119I, 2016MNRAS.456.1595M, 2018MNRAS.479.5060L, 2019NewAR..8601525D, 2019ApJ...877...17F, 2020ApJ...899..154S, 2020ApJ...892...29F, 2021ApJ...907...71F, 2021ApJ...922...83T}, though it is unclear if they will ever merge to form binaries.  Recently, the North American Nanohertz Observatory for Gravitational waves (NANOGrav) along with other pulsar timing arrays (PTAs) have determined that the common red noise process seen in measurements is caused by gravitational waves \citep{2023ApJ...951L...8A, 2023ApJ...951L...6R, 2023arXiv230616214A}.  It has been interpreted as the stochastic gravitational wave background (GWB) caused by MBHBs \citep{2023ApJ...952L..37A} or other exotic physics \citep{2023ApJ...951L..11A}. Given that MBHBs are thought to also dominate the formation of galaxy cores, it is possible to combine both electromagnetic and gravitational wave signals to lead to a better understanding of the MBHB population. This multimessanger approach is essential for astrophysical interpretations of the GWB, which has degeneracies that can be broken with EM observations \citep{2023ApJ...952L..37A}.

In this Paper we investigate how MBHB mass ratio distributions leave observable effects via the stellar cores they leave behind. In section \ref{OBS} we describe the methods used to calculate mass deficits and stellar masses from an observed sample of ETGs. Section \ref{TNG} details our post-processing of IllustrisTNG simulations to make predictions of mass deficits utilizing native cosmological merger trees. In section \ref{COM} we describe the statistical comparison between synthetic data sets generated from IllustrisTNG and the observed sample to obtain constraints on the MBHB mass ratios capable of creating a core and the maximum gas mass fraction allowed in a remnant before the core is erased. Lastly, we discuss our results in section \ref{DIS}. 

\section{Methods: Observational Data}
\label{OBS}

In this section we describe the the observed sample and how we derive the stellar mass of each galaxy and its mass deficit.

The upper limit for the angular size of a galaxy core in the local universe is on the order of a few arcseconds \citep{1997AJ....114.1771F}, requiring high spatial resolution imaging. As such, our parent sample consists of 219 local, ETGs with \emph{Hubble Space Telescope} (\emph{HST}) observations. This sample was compiled by \cite{2007ApJ...664..226L} using data from \cite{2001AJ....122..653R,2005AJ....129.2138L,2000ApJS..128...85Q,2003AJ....125..478L,1995AJ....110.2622L} and \cite{2001AJ....121.2431R}. Each galaxy within the sample has reported absolute V-band magnitudes, $M_V$, morphological types, and distances, along with derived Nuker-profile parameters. The S\'{e}rsic profile \citep{1963BAAA....6...41S} is typically used when information about the entire galaxy, including outer envelopes, is of interest. The Nuker profile \citep{1995AJ....110.2622L}, on the other hand, was developed with the inner galaxy in mind and best reveals inner cores. See \citet{2003AJ....125.2951G, 2004AJ....127.1917T, 2009ApJS..181..135H} for further discussion on the pros and cons of each profile. The Nuker profile has the form
\begin{equation}
    I(r)=2^{\frac{\beta - \gamma}{\alpha}}I_b\left(\frac{r}{r_b}\right)^{-\gamma}\left[1+\left(\frac{r}{r_b}\right)^{\alpha}\right]^{\frac{\gamma - \beta}{\alpha}},
	\label{eq:Nuker}
\end{equation}
where $r$ is the semi-major axis on the plane of the sky, $r_b$ is the break radius, $I_b$ is the surface brightness at the break radius, $\beta$ and $\gamma$ are the negative logarithmic surface brightness slopes at large and small radii respectively, and $\alpha$ determines the sharpness of the transition between inner and outer profiles.  The sample is split into three categories based on the inner slope $\gamma$ of the surface brightness profile. Galaxies for which $\gamma < 0.3$ are ``cores'', and galaxies that have $\gamma > 0.5$ are ``power-laws''. Galaxies with inner slope values between $0.3$ and $0.5$ are referred to as ``intermediates''.

Given a surface brightness profile, and assuming circular symmetry (which is satisfied for the inner portions of most galaxies) the total luminosity is found by integration:
\begin{equation}
    L=2\pi \int_{{\,0}}^{{\,\infty}}{{I(r)r\,dr}}.
    \label{eq:Luminosity}
\end{equation}
In order to obtain a luminosity deficit, one must know the un-cored profile of the galaxy. This is not an observable property, so instead an extrapolated profile must be used. The model we adopt is such that the inner slope $\gamma$ for a given profile is equal to the slope at the break radius. We adopt this local extrapolation over a global model to best capture conditions near the core. The extrapolated profile becomes
\begin{equation}
    I_\text{ex}(r)=I_b\left(\frac{r}{r_b}\right)^{(\gamma - \beta)/2}
    \label{eq:ExNuker}
\end{equation}
for regions $r\in (0,r_\text{b}]$. Regions beyond the break radius continue to be described by the full Nuker profile. We then define the luminosity deficit as the difference between the integrated surface brightness profiles
\begin{equation}
    \Delta L=2\pi \int_{{\,0}}^{{\,r_b}}{{[I_\text{ex}(r) - I(r)]r\,dr}}.
    \label{eq:LumDef}
\end{equation}
The surface brightness values provided are given in $\text{mag}$ $\text{arcsec}^{-2}$, which we convert to $L_{\odot}$ pc$^{-2}$ via the usual:
\begin{equation}
    I\left[\frac{L_{\odot}}{\text{pc}^2}\right]=4.255\times10^8\cdot 10^{0.4(V_{\odot}-I\left[\frac{\text{mag}}{\text{arcsec}^2}\right])},
    \label{eq:IntensityConversion}
\end{equation}
where $V_\odot$ is the absolute magnitude of the sun in the V-band.

\begin{figure}
\includegraphics[width=\columnwidth]{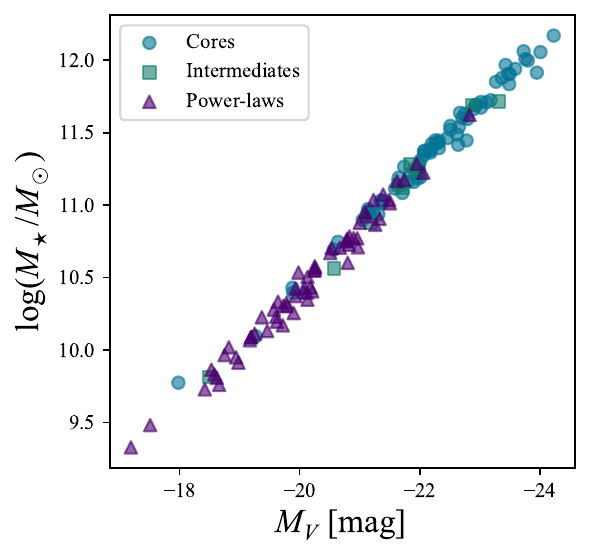}
    \caption{The stellar mass of each galaxy was calculated with equation (\ref{eq:Upsilon}). Here the values are plotted against the reported absolute V-band magnitude. The sample spans over 2 dex in stellar mass, allowing good probes of any trends in $M_\star$}
    \label{fig:M_Mv}
\end{figure}

\begin{figure}
	\includegraphics[width=\columnwidth]{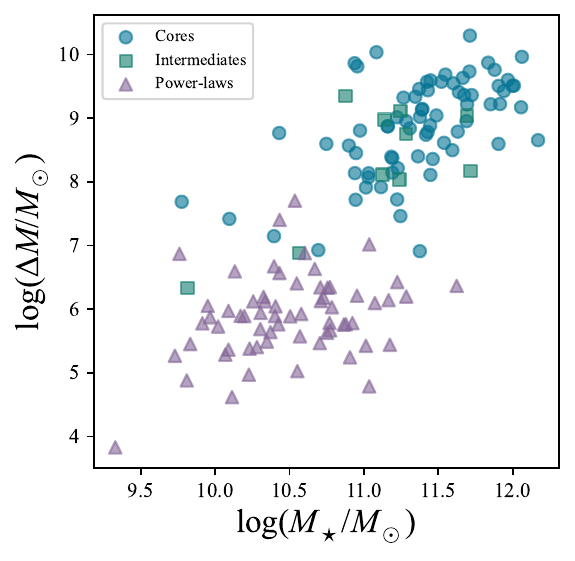}
    \caption{Relationship between the stellar mass deficit and stellar mass. $\Delta M$ is calculated as outlined in section 2 and shows a positive correlation with $M_\star$. The bimodality of cores and power-laws can be seen in the data, with cores tending to be more massive with higher mass deficits.}
    \label{fig:dM_M}
\end{figure}

In order to produce a mass deficit from the luminosity deficit a mass-to-light ratio, $\Upsilon$, for each galaxy is required. We use the $\Upsilon$--color relation developed by \citet{2003ApJS..149..289B}:
\begin{equation}
    \log_{10}(\Upsilon_V)=a_V + (b_V \cdot \text{color})
    \label{eq:Upsilon},
\end{equation}
such that $a_V=-0.628$ and $b_V=1.305$. For optical $\Upsilon$ the scatter is $\sim 0.1$ dex. Calculated values of $\Upsilon_V$ are also used to determine the stellar mass of each galaxy from their V-band luminosities. A total of 149 galaxies in our sample had extinction corrected $(B-V)$ indices available in the HyperLeda\footnote{http://leda.univ-lyon1.fr/} database \citep{2014A&A...570A..13M}, which we used to compute $\Upsilon_V$.
We calculated luminosity deficits for each galaxy numerically using equation (\ref{eq:LumDef}) and converted them into stellar mass deficits using the mass-to-light ratio ($\Delta M=\Upsilon_V \Delta L_V$). Figure \ref{fig:M_Mv} shows the results of the stellar mass calculations plotted against the reported V-band absolute magnitude, and Figure \ref{fig:dM_M} shows the calculated mass deficit versus stellar mass. The bimodality of cores and power-laws is readily apparent, with cores being more massive and having higher mass deficits \citep{2007ApJ...664..226L}.

\begin{figure*}
	\includegraphics[width=\textwidth, keepaspectratio]{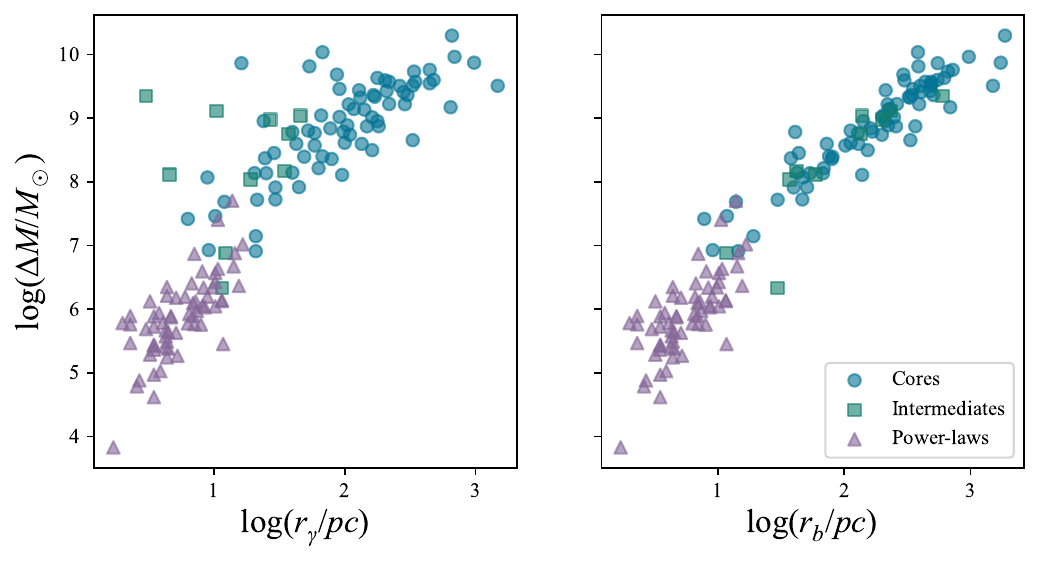}
    \caption{The stellar mass deficit $\Delta M$ is expected to be a function of the cusp radius $r_\gamma$ or the break radius $r_b$ \citep{2007ApJ...662..808L}. The figure shows the calculated values of the mass deficit to be consistent with this relationship, with $\Delta M$ and $r_b$ showing less scatter. This is expected, as our prescription for mass deficit depends on $r_b$}
    \label{fig:dM_r}
\end{figure*}

\begin{figure}
    \includegraphics[width=\columnwidth]{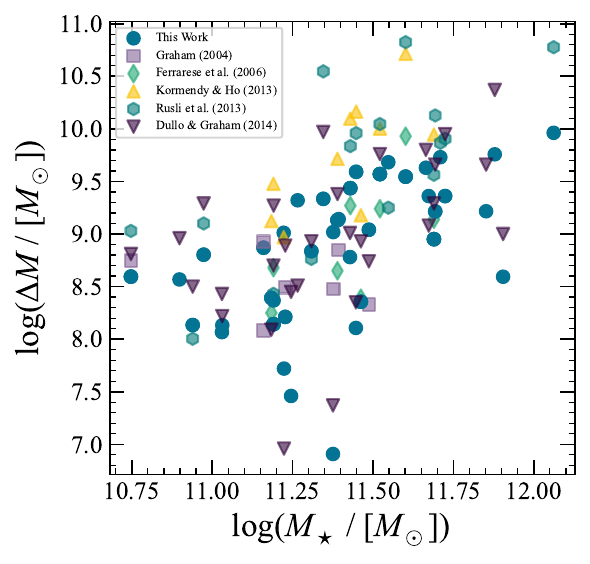}
    \caption{Comparison of mass deficits for our sample galaxies which overlap with those observed in other studies. Mass deficits calculated in this work are marked as blue circles while those of \citet{2004ApJ...613L..33G} are lavender squares, \citet{2006ApJS..164..334F} are green diamonds, \citet{2013ARA&A..51..511K} are yellow triangles, \citet{2013AJ....146..160R} are blue-green hexagons, and \citet{2014MNRAS.444.2700D} are purple inverted triangles. Mass deficits of other works were taken from tables or else calculated using methods described in the respective papers. Our $\Delta M$--$M_\star$ trend is consistent with previous literature, except in the cases of \citet{2013ARA&A..51..511K} and \citet{2013AJ....146..160R} where our values tend to be an order of magnitude lower. Disparate values are due to differing model choices, including adopted mass-to-light ratios, surface brightness profiles (Core-S\'{e}rsic, S\'{e}rsic, or Nuker), and outer profile extrapolations.}
    \label{fig:Lit_comp}
\end{figure}

If MBHBs are responsible for core souring, it is expected that the mass in ejected stars should positively correlate with the cusp radius $r_\gamma$ (or the break radius $r_b$) \citep{2007ApJ...662..808L}. The cusp radius is related to the break radius by
\begin{equation}
    r_\gamma \equiv r_b \left(\frac{\gamma'-\gamma}{\beta - \gamma'}\right)^{1/\alpha},
    \label{eq:cuspradius}
\end{equation}
where $\gamma'$ is defined as
\begin{equation}
    \gamma'(r_0)\equiv - \left.\frac{d\mathrm{log}I}{d\mathrm{log}r}\right|_{r=r_0}
    \label{eq:gammap}
\end{equation}
with $r_0$, in this case, the \textit{HST} resolution scale. In principle $r_0$ can be any limiting radius. Figure \ref{fig:dM_r} shows the expected relationship between mass deficit and both the cusp and break radii, demonstrating that the numerical prescription used here to calculate mass deficits is consistent with expectations. The mass deficit values are also reasonable, with the highest values belonging to the brightest cluster galaxies (BCGs), which have $\Delta M \sim 10^{10} \ M_\odot$. Further, our mass deficit values show a trend consistent with those calculated from Core-S\'{e}rsic models \citep[e.g.,][]{2004ApJ...613L..33G, 2006ApJS..164..334F, 2014MNRAS.444.2700D}, or else our values tend to be about an order of magnitude lower \citep[e.g.,][]{2009ApJS..182..216K, 2013ARA&A..51..511K, 2013AJ....146..160R}, which Figure \ref{fig:Lit_comp} demonstrates. Disparate values are to be expected due to the sensitivity of observed mass deficits to model choices, including the adopted mass-to-light ratios, surface brightness profile, and how one extrapolates the outer profile \citep{2010MNRAS.407..447H, 2021ApJ...922...40D}. \cite{2013ARA&A..51..511K}, for example, use mass-to-light ratios $\Upsilon_V\propto L_V^{0.36}$ which end up being a factor of $\sim 3\times$ higher than our values. Additionally, \cite{2013ARA&A..51..511K} use a global S\'{e}rsic fit and extrapolate inwards, resulting in a sensitivity to larger-scale deviations. In this work we extrapolate using a local Nuker fit, making our model more sensitive to local values.

\begin{figure}
	\includegraphics[width=\columnwidth]{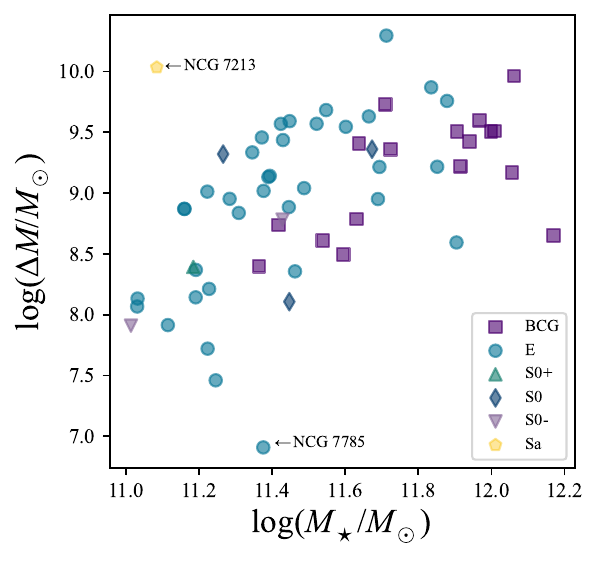}
    \caption{Mass deficit versus stellar mass for core galaxies with $M_\star \gtrsim 10^{11} \ M_\odot$ colored by morphological type. The BGC galaxies appear to fall below the trend suggested by the elliptical and lenticular galaxies. NGC 7213 is an Sa galaxy, while NGC 7785 appears to be an outlier. Both galaxies are removed from the sample when compared with TNG300 data.}
    \label{fig:dM_M_c}
\end{figure}

Of the 149 ETG galaxies for which we have mass deficits, 60 are cores. These are plotted in Figure \ref{fig:dM_M_c} and colored by morphology. The BCGs, colored in purple, seem to fall below the trend suggested by the elliptical and lenticular galaxies. This could be explained by the fact that BCGs are not statistical, being heavily dominated by past mergers \citep{2007ApJ...662..808L}, for which the linear relationship between number of dry mergers and stellar mass deficit breaks \citep{2006ApJ...648..976M}. Of note, we remove the Sa (non-ETG) galaxy NGC 7213 and NGC 7785 which appears to be an outlier, resulting in 58 core galaxies used in our final analysis.

\section{Methods: Illustris TNG300-3}
\label{TNG}

In this Paper we use IllustrisTNG as an avenue for constraining the minimum mass ratio of MBHBs capable of scouring a core, as well as the maximum gas mass fraction, above which a core will be erased. The IllustrisTNG project is a cosmological magnetohydrodynamical simulation suite for galaxy formation, spanning a range of volume and resolution. Its stated purpose is to elucidate when and how galaxies evolve into the structures observed at $z=0$. A major component of the simulation is the SubLink algorithm \citep{2015MNRAS.449...49R} used to construct merger trees at the subhalo level, where ``subhalo'' is interchangeable with ``galaxy''. The merger trees for each galaxy can be traced back, where information about each progenitor can be obtained, including the mass ratio of the subhalos involved in a merger event as well as their gas mass.
We use the results of the publicly available TNG300-3 simulation volume and resolution, as described in \cite{2019ComAC...6....2N}, to make predictions about how MBHB properties influence the creation of galaxy cores. We selected the 300 Mpc volume since it gives the best properties and statistics for the massive galaxies of interest, including the stellar mass (which informs SMBH mass) and merger history of each galaxy. Since mass assembly for galaxies with stellar masses $M_\star \gtrsim 10^{11} M_\odot$ is dominated by mergers, we expect core scouring to be most prominent in this mass regime. We take all 1898 subhalos above this lower limit as our simulation sample.

IllustrisTNG does not possess the resolution scale required to resolve core structure. The softening length in TNG300-3 for collisionless particles (dark matter and stars) is $\epsilon_{\text{DM},\star}(z=0) = 5.90$ kpc \citep{2017MNRAS.465.3291W, 2018MNRAS.473.4077P}, while the range of break radii in our core galaxy sample is $11.8 \text{ pc} < r_b < 1870 \text{ pc}$. Accordingly, we perform sub-grid post-processing prescriptions to calculate stellar mass deficits. We use the results of N-body simulations presented by \cite{2006ApJ...648..976M} to calculate the expected mass deficit for each subhalo. In these simulations a primary SMBH is initially located at the center of a galaxy, and a secondary SMBH is allowed to spiral in. Early in the simulation the orbit of the secondary decays due to DF and at later times the system evolves via both DF and stellar scattering. During the latter phase, the mass in ejected stars is tracked and constitutes the mass deficit. Five binary mass ratios were considered, with values $q = (0.5, 0.25, 0.1, 0.05, \textrm{and}\ 0.025)$. For each value of $q$ three different density profiles were explored, where the inner density slopes were $\gamma=d\log{\rho}/d\log{r} = (0.5, 1.0, \textrm{and}\ 1.5)$. These models were integrated with two different values of $N$: $1.2\times10^{5}$ and $2.0\times10^{5}$ particles. It was found that the mass in ejected stars follows the linear relation
\begin{equation}
    \Delta M = f \mathcal{N} M_\text{BH},
    \label{eq: massdef}
\end{equation}
where $f$ is a scaling factor, $\mathcal{N}$ is the number of mergers contributing to core scouring and $M_\text{BH}$ is the mass of the remnant black hole. We use this relation along with merger histories provided by the TNG300 merger trees to calculate the mass deficit.

Observations have shown a range of values $0.5 \lesssim \Delta M/M_\text{BH} \lesssim 10$ \citep[e.g.,][]{2006ApJS..164..334F, 2013ARA&A..51..511K, 2013AJ....146..160R, 2014MNRAS.444.2700D}. Thus the product $f\mathcal{N}$ should also fall in this range.  Theoretical estimates of $f$ range between 0.5 for spherical systems \citep{2006ApJ...648..976M} and 5 for triaxial systems, though the ejected mass in triaxial systems does not solely come from the innermost regions \citep{2012ApJ...749..147K}, a point we return to in section \ref{DIS}.  We determine the number of past mergers $\mathcal{N}$ a subhalo has had by following the main (primary) progenitor branch back in time as well as following the branches of the next (secondary) progenitors to obtain the full progenitor history. This is done starting at snapshot 100 (corresponding to $z\approx0$ and continuing back to snapshot 33 (mapping to $z\approx2$). Mergers are counted only when a minimum mass ratio $q$ and a maximum gas mass fraction $f_\text{gas}$ are satisfied.  The values are set by our fitting procedure described in section \ref{COM}. Establishing a mass ratio cut-off reflects the minimum MBHB mass ratio required to successfully scour out a core, while a gas mass fraction constraint models the condition for core erasure. The mass ratio is defied to be
\begin{equation}
    q\equiv \frac{M_{\star,2}}{M_{\star,1}},
\end{equation}
where $M_{\star,2}$ is the stellar mass of the secondary subhalo within the half mass radius, and similarly for the primary subhalo mass $M_{\star,1}$. The gas mass fraction is
\begin{equation}
    f_\text{gas}\equiv \frac{M_{g,1}+M_{g,2}}{M_{\star,1}+M_{\star,2}+M_{g,1}+M_{g,2}},
\end{equation}
where $M_{g,1}$ and $M_{g,2}$ are the gas mass within the half-mass radius of the primary and secondary respectively. A merger resulting in a remnant which exceeds a provided gas mass fraction is considered a gas-rich (wet) merger while values less than the limit count as gas-poor (dry) mergers. We assume that once a wet merger has occurred any core formation will be erased by subsequent star formation. Thus only the most recent, consecutive dry mergers count toward calculating mass deficits. Since $\mathcal{N}$ is determined by $q$ and $f_\text{gas}$, $f$ is a free parameter.

The mass of the merged black hole is determined using the $M_\text{BH}$--$M_{\text{bulge}}$ relation \citep{2013ARA&A..51..511K},
\begin{equation}
    \frac{M_\text{BH}}{10^9 \ M_\odot}=\left(0.49_{-0.05}^{+0.06}\right)\left(\frac{M_{\text{bulge}}}{10^{11} \ M_\odot}\right)^{1.16\pm0.08}; \sigma=0.29 \ \text{dex},
    \label{eq:Mbh-Mbulge}
\end{equation}
assuming for our mass range all subhalos will be bulge dominated. When calculating the mass deficit using equation \ref{eq: massdef} we introduce the scatter of 0.29 dex to determine $M_\text{BH}$ for each subhalo. Since equation \ref{eq:Mbh-Mbulge} shows a power-law index of about unity, $q$ is approximately the BH mass ratio.

\begin{figure*}
    \includegraphics[width=\textwidth, keepaspectratio]{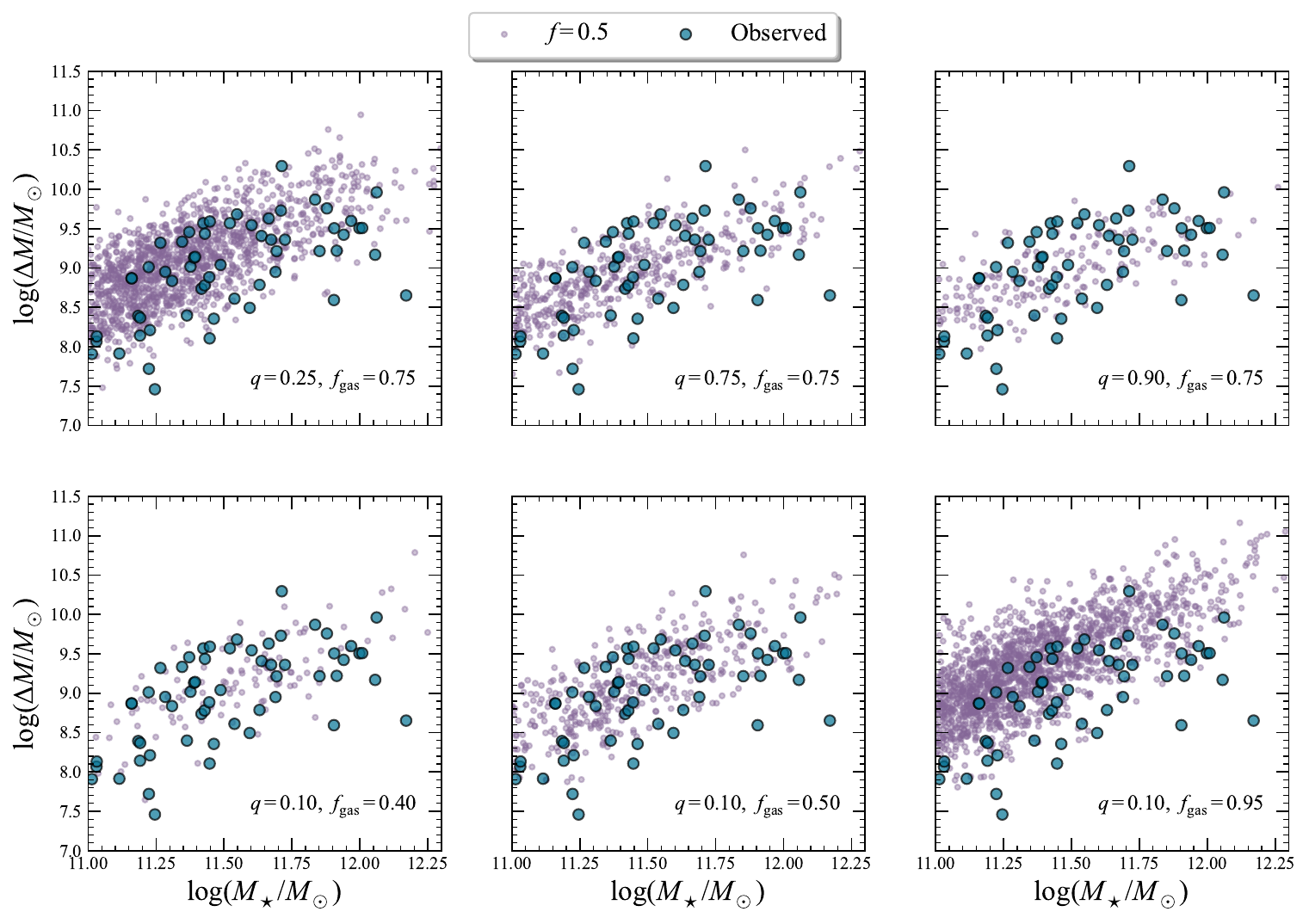}
    \caption{The grid of plots shows mass deficit versus stellar mass for both the observed sample (large blue circles) and synthetic data sets (small purple circles) generated from combinations $q$ and $f_\text{gas}$, determining $\mathcal{N}$, and the factor $f=0.5$ in equation \ref{eq: massdef}. The top row holds the maximum gas mass fraction required to produce a core fixed while the minimum mass ratio is allowed to vary, while in the bottom row the reverse is true. A low mass ratio threshold results in more core galaxies compared to high thresholds when keeping $f_\text{gas}$ fixed. A high gas mass fraction threshold, for fixed $q$, results in more core galaxies. Additionally, for low $q$ and high $f_\text{gas}$, mass deficits are higher on average than for other combinations.}
    \label{fig:grid}
\end{figure*}

\begin{figure}
    \includegraphics[width=\columnwidth]{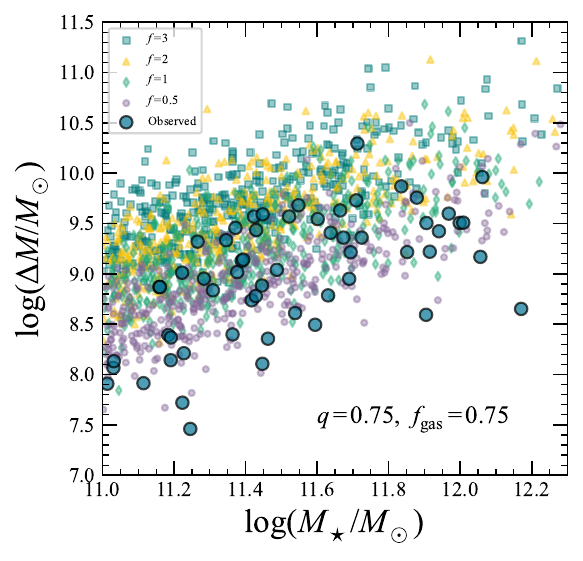}
    \caption{Mass deficit versus stellar mass for both the observed sample (large blue circles) and synthetic data sets generated from $q=f_\text{gas}=0.75$, determining the value of $\mathcal{N}$, and the factor $f$ in equation \ref{eq: massdef} Synthetic data sets are marked according to the value of $f$ chosen, with $f=0.5$ corresponding to purple circles, $f=1$ to green diamonds, $f=2$ to yellow triangles, and $f=3$ to light blue squares. Increasing $f$ can be seen to simply increase the average mass deficit at fixed $M_\star$.}
    \label{fig:MdM_All}
\end{figure}

It is expected that a low mass ratio threshold will result in a larger population of core galaxies. Physically, this is a statement about the efficiency of core scouring. Setting a low $q$ means that smaller secondary black holes do make it to the bottom of the potential well and that the binary contributes to stellar scattering. Similarly, a high gas mass fraction threshold for what counts as a gas-poor merger increases the chances a core will not be erased, according to our prescription. Physically, this means that the gas funneled to the nucleus does not result in star formation. If stellar scattering is more efficient and star formation less efficient, then the average mass deficit across the population of core galaxies will \emph{increase}. At the same time, inefficient scattering and efficient nuclear star formation must result in a lower mass deficit in the core population. Figures \ref{fig:grid} \& \ref{fig:MdM_All} show different realizations of calculated mass deficits for TNG300 subhalos using combinations of $f$, $q$, and $f_\text{gas}$ to illustrate the above statements.

\section{Comparison of Simulation to Observations}
\label{COM}
Here we compare the parameterized model of cores created by MBHBs with the observed data to determine which mass ratio binaries are capable of core scouring and what minimum gas mass fraction leads to core erasure. Once a synthetic data set is created using our model we could directly compare it to a volume-limited sample of core galaxies, but our sample is heterogeneous. Thus we require an extra step to calculate the expected mass deficit given a galaxy stellar mass.

The observed data are compared to the TNG300 subhalos using the \texttt{emcee} Python MCMC package \citep{2013PASP..125..306F}. The likelihood function is calculated for a given data point, $i$, with stellar mass and mass deficit values $M_{\star,i},\ \Delta M_i$ as
\begin{equation}
    \mathcal{L}_i(M_{\star,i},\Delta M_i|\vec{\theta})=\frac{K(M_{\star,i},\Delta M_i)}{\int_0^\infty K(M_{\star,i},\Delta M_i) \ d\Delta M_i}.
\end{equation}
Here, $\vec{\theta}=[q, \ f_\text{gas}]$ are our parameters of interest. The function $K(M_{\star,i},\Delta M_i)$ is a kernel density estimation (KDE) evaluated at the data point. The KDE is calculated with standard routines, taking as an argument a synthetic sample of TNG300 subhalos generated as described in section \ref{TNG} using the specified combination of $q$ and $f_\text{gas}$. To ensure a smooth KDE a scatter of 0.2 dex is introduced to the stellar mass before the mass deficit is calculated.  This method is iterated until the synthetic sample has $\ge 10000$ data points. Combinations of $q$ and $f_\text{gas}$ which result in fewer than 50 core galaxies before iteration are ignored as they are not capable of producing the observed prevalence of core galaxies. The function $K$ is normalized by its integral over all $\Delta M$ in order to penalize models that predict values of $\Delta M_i$ that are too high or low for a given value of $M_{\star, i}$.  This avoids problems with an unnormalized $K$, which would penalize models that do not predict distributions that are as clustered as the heterogeneous sample. Our observational sample is not a volume-limited sample, and selection effects would severely bias direct comparison in the two-dimensional space without instead calculating the expected distribution of $\Delta M$ for a fixed value of $M_{\star}$, as we do here.

\begin{figure}
\centering
	\includegraphics[scale=0.64]{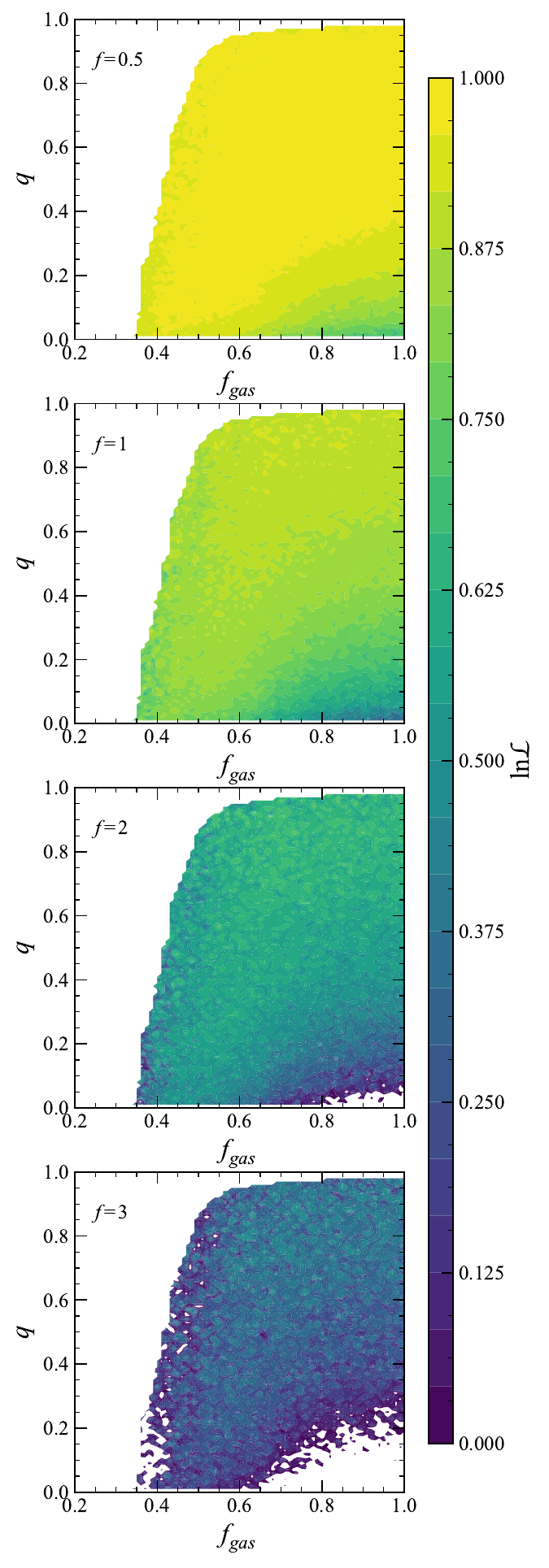}
    \caption{Log-likelihood calculations with a 100 $\times$ 100 grid of $q$ and $f_\text{gas}$ values with a range of [0.01, 1]. The four panels represent different calculations using $f=0.5, 1, 2, 3$, starting from the top panel. The log-likelihood is normalized to the maximum value found among all calculations. The scenario where $f=0.5$ has the highest log-likelihoods, despite the stochastic nature of the calculations.}
    \label{fig:Like_Grid}
\end{figure}

Figure \ref{fig:Like_Grid} shows 100 $\times$ 100 grid calculations of the log-likelihood as a function of $q$ $\in$ [0.01, 1] and $f_\text{gas}$ $\in$ [0.01, 1] for a range of values $f = 0.5, 1, 2, 3$, normalized to the maximum value among all calculations. The case where $f=0.5$ has the highest log-likelihoods, obvious despite the stochastic nature of the calculations. We therefore take this this to be the constant value of $f$ when conducting MCMC fitting.

\begin{figure}
    \includegraphics[width=\columnwidth]{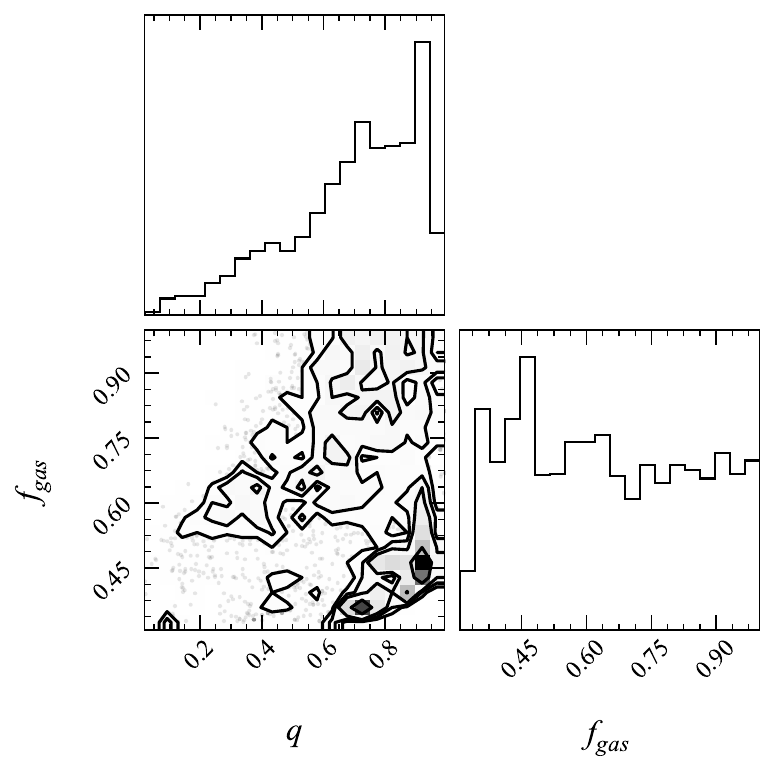}
    \caption{Corner plot for the best fit minimum mass ratio and maximum gas mass fraction for the observed data. The histogram for $q$ is unimodal, with a sharp peak and rapid fall-off towards low mass ratios. The model rules out mass ratios $q < 0.25$ at 95$\%$ confidence. On the other hand, the histogram for $f_\text{gas}$ shows a peak near 0.45, though the distribution is broad.}
    \label{fig:TNGCorner}
\end{figure}

\begin{figure}
    \includegraphics[width=\columnwidth]{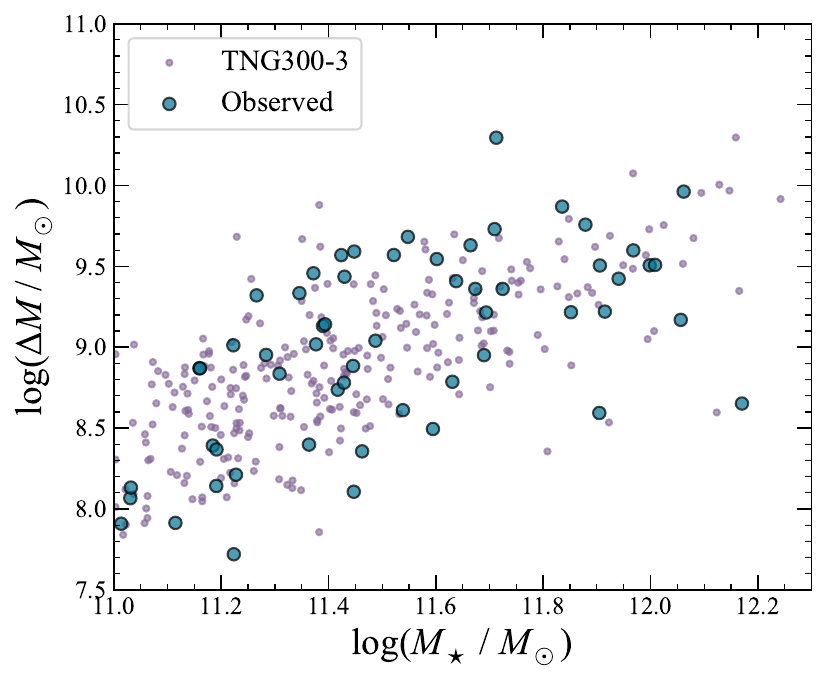}
    \caption{A comparison between the observed mass deficit as a function of stellar mass and a sample synthesised from TNG300-3 using the median values for the $q$ and $f_\text{gas}$ posterior samples. The main trend in the observed data is recovered. Note that the absolute number of core galaxies is not a constraint, as our sample is heterogeneous. It is only the main trend and overall values of $\Delta M$ for a given $M_\star$ we are able to reproduce.}
    \label{fig:compared}
\end{figure}

The MCMC algorithm is provided the log-likelihood ($\ln{\mathcal{L}} = \sum_i \ln{\mathcal{L}_i}$) and explores the parameter space using flat priors on both the minimum mass ratio and maximum gas mas fraction. The bounds on these parameters are taken to be [0.01, 1.00]. Figure \ref{fig:TNGCorner} shows the results of the fitting process. The median values of the posterior samples and 68\% confidence intervals are $q=0.7^{+0.2}_{-0.3}$ and $f_\text{gas}=0.6^{+0.3}_{-0.2}$. The distribution of $q$ values is unimodal and sharply peaked with a rapid fall-off towards small values. Mass ratios of $q< 0.25$ are ruled out at the 95$\%$ confidence level. The maximum gas mass fraction distribution is broader, but shows a clear peak near 0.45. Figure \ref{fig:compared} shows a simulated sample with calculated mass deficits using the median values of the posterior distributions of $q$ and $f_\text{gas}$. The observed sample is over-plotted for visual comparison. We find the predicted trend to be in good agreement with observation.

\section{Discussion}
\label{DIS}
The results of the fitting procedure show that the minimum mass ratio for core formation is well constrained in our model. Since $M_\text{BH}\sim M_\star^\alpha$ with $\alpha$ approximately unity (see equation \ref{eq:Mbh-Mbulge}), this is a direct constraint on the MBHB mass ratio. Low mass ratio binaries, therefore, do not contribute to core scouring. Further, up to around half the mass within the half-mass radius of a remnant ETG can consist of gas without the core being erased. Gas content much larger than this would predict larger mass deficits than we observe (as shown in Figure \ref{fig:grid}).

Throughout the analysis we have made a number of practical assumptions to make comparisons between observational data and synthetic data.  In particular, we have assumed that (1) the extrapolated surface brightness profile of each galaxy is a simple power-law (2) a single gas-rich merger is capable of erasing all core structure in a galaxy, (3) all MBHBs coalesce before a subsequent merger (i.e., there are never any triple BH systems), (4) the eccentricity of the MBHB orbit is irrelevant, (5) the mass ejected in stars is always a linear function of $\mathcal{N}$, (6) the $M_\text{BH}-M_\text{bulge}$ relation does not evolve with redshift, and (7) DF and stellar scattering timescales are irrelevant. We consider below the effects of the above assumptions on our results.

For any single galaxy, missing-light measurements are sensitive to the selected model \citep{2010MNRAS.407..447H, 2021ApJ...922...40D}, though measurements over a population of galaxies should be largely accurate, making our assumption of a power-law, un-cored profile a fair approximation. 
As mentioned above, the total mass ejected is a simple product of $\mathcal{N}$, the number of mergers, and $f M_{\mathrm{BH}}$, an amount of mass scattered per merger that scales with mass of the black hole.  For most of our calculations, we assume $f = 0.5$, but there are simulation-based estimates for $f$ as large as 5 \citep{2012ApJ...749..147K}.  The value for $\mathcal{N}$ comes from the TNG300-3 simulations, whose merger tree is consistent with observed pair-fractions and merger rates \citep{2004ApJ...608..752B, 2012ApJ...744...85M, 2015MNRAS.449...49R}.  We also find the best agreement between theory and observations for $f = 0.5$, though our preferred values of $q$ and $f_{\mathrm{gas}}$ do not change (Figure\ \ref{fig:Like_Grid}).

It is worth understanding why the more realistic triaxial simulations due to \citet{2012ApJ...749..147K} result in values of $f > 1$ whereas the spherical simulations due to \citet{2006ApJ...648..976M} predict $f = 0.5$ in best agreement with our data.
The triaxial systems include orbits that come from large radii to refill the loss cone, thus enabling the binary to merge within a Hubble time \citep{2002MNRAS.331..935Y, 2006astro.ph..1520H, 2012ApJ...749..147K, 2015ApJ...810..139H, 2023MNRAS.524.4062M}.  Because stars on these orbits are responsible for the removal of a substantial amount of energy from the binary, they constitute a significant amount of the total mass deficit.  These stars, however, likely do not contribute a substantial amount of light at the center of the galaxy. Thus while the total amount of mass ejected is larger (as is necessary to solve the final parsec problem), the amount of observable mass deficit is  much smaller and likely to be well approximated by the amount of mass ejected in a spherical galaxy, since this will only remove stars from the central part of the galaxy where light deficits are observed.

Assuming a single wet merger can completely erase core structure will tend to underestimate the final size of the core, and therefore the stellar mass deficit. It may be an oversimplification that all gas-rich mergers result in full core erasure, since not all gas will be cold, star-forming gas and the central SMBH of a remnant can act as a guardian, protecting the core \citep{1997AJ....114.1771F}.

N-body simulations would be needed to understand the effect of a triple BH system on stellar scattering, including the possibility of the ejection of the primary and secondary from the system \citep[e.g.,][]{2007MNRAS.377..957H, 2008ApJ...678..780G, 2023MNRAS.524.4062M}. A tertiary black hole can possibly decrease the mass deficit by stalling stellar scattering, extracting energy from the primary and secondary, hardening the orbit and ``skipping'' a large portion of the scattering phase \citep[e.g.,][]{2002ApJ...578..775B}. A highly eccentric MBHB can also decrease the mass deficit, as binaries with large eccentricities coalesce on a shorter timescale \citep{1963PhRv..131..435P, 1964PhRv..136.1224P}. Thus the size of cores also depends on the distribution of MBHB eccentricities.

\begin{figure}
	\includegraphics[width=\columnwidth]{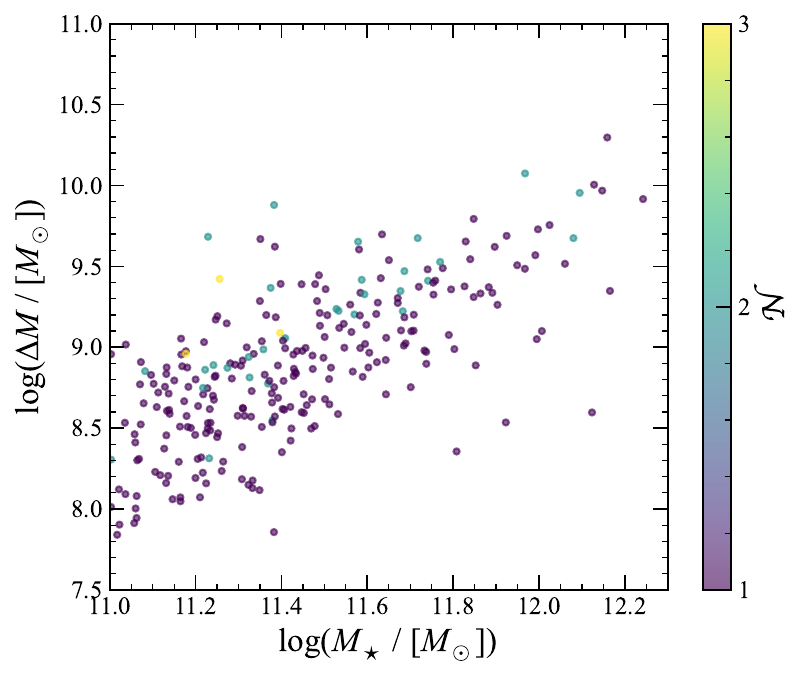}
    \caption{
    Mass deficit versus stellar mass for for the set of TNG300-3 subhalos 
    designated as core galaxies when the minimum mass ratio for core 
    scouring and the maximum gas mass fraction are set to the median values obtained from MCMC fitting. Each subhalo is colored according to the number of most recent, 
    consecutive gas-poor mergers it has had. There are no subhalos with 
    $\mathcal{N} > 3$ in this simulated sample, which is consistent with the results described in 
    \protect{\citet{2006ApJ...648..976M}} and with hierarchical structure formation \citep[e.g.,][]{2004ApJ...608..752B, 2012ApJ...744...85M}.}
    \label{fig:dM_M_merg}
\end{figure}

Assuming a linear relation between number of mergers and mass deficit may be an oversimplification, as shown by \cite{2006ApJ...648..976M}. In N-body experiments the the mass deficit as a function of $\mathcal{N}$ is no longer linear after $\gtrsim 3$ dry merger events. However, Figure \ref{fig:dM_M_merg} shows there are no galaxies in the synthetic sample whose number of recent, consecutive gas-poor mergers exceeds three. This is consistent with the results described in \protect{\citet{2006ApJ...648..976M}} and with hierarchical structure formation \citep[e.g.,][]{2004ApJ...608..752B, 2012ApJ...744...85M}.

If the $M_\text{BH}$--$M_\text{bulge}$ relation has a larger normalization at higher redshifts, then we might expect larger cores. MBHB binaries formed at higher redshifts would have larger masses compared to their host galaxy's total stellar mass, leading to more stellar scattering. Such an evolution would imply larger values of $q$ and/or smaller values of $f_\text{gas}$. 

The minimum mass ratio of MBHBs capable of carving out a core and the maximum gas content are affected by the assumption DF and stellar scattering timescales can be neglected. We have not considered in our treatment how the effective radius, central velocity dispersion, or the nuclear stellar density of the progenitor galaxies contribute to whether the secondary BH forms a binary with the primary. In future work these physical considerations can be added to the post-processing phase when determining predicted mass deficits and thus lead to further constraints on MBHB properties. An additional constraining factor is the number of core galaxies observed within some volume. The current sample is not volume-limited and is therefore not a representative population of core galaxies, though a statistical comparison between the number of core galaxies observed and those predicted by IllustrisTNG would provide an important constraint. This can be achieved in the future by creating a volume-limited sample using archival \textit{HST} data or new observations.

Even when making these assumptions, we are able to see a clear picture emerge in which (i) we can rule out low-mass mergers as relevant for core scouring and (ii) we find gas-rich mergers must be capable of erasing cores or else we would find larger mass deficits than are seen in the data.

\section*{Summary}
\label{summary}
In this Paper we presented a general method for computing stellar mass deficits, which involved parameterization of the light distribution of a galaxy in question and the adoption of a model for the original central light profile.  Once we derived a surface brightness profile, and selected a model, the light deficit was computed by taking the integrated difference between them (see Figure \ref{fig:MassDefEx}). The missing light was then translated into a stellar mass deficit by the multiplication of a mass-to-light ratio. We followed this prescription for galaxies with fitted Nuker profiles in section 2.

Although the model of core scouring has been well developed, the connection between the properties of core galaxies and the massive black hole binary population is poorly constrained. A major reason for this is that the merger history of any individual galaxy is largely unknown. We made progress in solving this problem by using TNG300-3 to obtain subhalo stellar masses and and merger histories via SubLink merger trees. We applied sub-grid, post-processing physics (described in section 3) to create synthetic $\Delta M$--$M_\star$ trends to compare with the observed trend using MCMC methods (section 4).

We found the median values of the posterior distribution for our parameters to be $q=0.7^{+0.2}_{-0.3}$ and $f_\text{gas}=0.6^{+0.3}_{-0.2}$. Low mass ratio binaries do not contribute to core formation, and gas-rich mergers must be capable of erasing cores. These constraints add to our understanding of the distribution of MBHB mass ratios and the expected rates of these binaries. Further, the distribution of mass ratios for MBHBs contributing to core scouring is consistent with constraints on the MBHB determined using NANOGrav's 15-year PTA data set \citep{2023ApJ...952L..37A}. These results and future endeavours to increase understanding of the MBHB demographic will affect how the origins of the GWB are approached along with future science done by PTAs like NANOGrav and in the course of the \emph{LISA} mission \citep{2023arXiv230113854E, 2023LRR....26....2A}.

%%%%%%%%%%%%%%%%%%%%%%%%%%%%%%%%%%%%%%%%%%%%%%%%%%
\section*{Acknowledgements}
\label{Acknowledgements}

The Authors would like to thank Tod Lauer and Luke Kelly for helpful discussions. We would also like to thank the referee for his/her/their valuable comments, which have greatly improved the Paper.

%%%%%%%%%%%%%%%%%%%%%%%%%%%%%%%%%%%%%%%%%%%%%%%%%%
\section*{Data Availability}

The parent sample data underlying this article are available at https://dx.doi.org/10.26093/cds/vizier.16640226. Data generated in this work will be made available for download.
 
%The inclusion of a Data Availability Statement is a requirement for articles published in MNRAS. Data Availability Statements provide a standardised format for readers to understand the availability of data underlying the research results described in the article. The statement may refer to original data generated in the course of the study or to third-party data analysed in the article. The statement should describe and provide means of access, where possible, by linking to the data or providing the required accession numbers for the relevant databases or DOIs.



\bibliographystyle{mnras}
\bibliography{example} % if your bibtex file is called example.bib

%\begin{thebibliography}{99}
%\end{thebibliography}

%%%%%%%%%%%%%%%%%%%%%%%%%%%%%%%%%%%%%%%%%%%%%%%%%%

%%%%%%%%%%%%%%%%% APPENDICES %%%%%%%%%%%%%%%%%%%%%

% \appendix

% \section{Some extra material}


%%%%%%%%%%%%%%%%%%%%%%%%%%%%%%%%%%%%%%%%%%%%%%%%%%

% Don't change these lines
\bsp	% typesetting comment
\label{lastpage}
\end{document}